# ALGORITHMS OF EVALUATION OF THE WAITING TIME AND THE MODELLING OF THE TERMINAL ACTIVITY BASED ON THE TRAFFIC COEFFICIENT IN THE SEAPORT

*Gh. Mişcoi, A. Costea, I.R. Ţicu, C. Pomazan*

**Abstract**
This paper approaches the application of the waiting model with Poisson inputs and priorities in the port activity. The arrival of ships in the maritime terminal is numerically modelled, and specific parameters for the distribution functions of service and of inputs are determined, in order to establish the waiting time of ships in the seaport and a stationary process. The modelling is based on waiting times and on the traffic coefficient.

**1. Introduction**

It is known that both the arrivals of ships in the sea port, as well as the operations upon them (loading, unloading, etc.) can be effectively described and investigated by using waiting models (see, for example monograph [1]). It is also demonstrated that service laws primarily appear as optimal in the class of all service laws [2]. The present work is searching for a waiting model with Poisson inputs, ordered in five priority classes, with comprehensive service and arbitrary service distribution functions. Generally, (for an arbitrary number of priority classes) this model, briefly denoted by the abbreviation $M_r | G_r |1$ is described and investigated in monographs [3, 4]. In the books mentioned, the main performance characteristics of the model evolution, including the average waiting time, the stationary conditions and the traffic coefficient are obtained. However, the average waiting time, the traffic coefficient and the stationary conditions are expressed as a rule, through the Laplace-Stieltjes transforms of the distribution functions of service. The problem is that for shaping these features, we must find the numerical values of the Laplace-Stieltjes transforms for certain values of the parameter of the summary flow. In models with priority, several laws of priority are used. In this paper, tables showing the dependence of the waiting time in relation to various parameters and tables with numerical modelling of the traffic coefficient for three strategies of service with absolute priority are presented: 1) the case of continuing the discontinued service; 2) the case of losing the service and 3) the case of serving from the beginning an interrupted service.

Next, we shall introduce the following denotations. We denote the parameter of the Poisson input stream by $a$; $B_k(x)$, the distribution function of service for ships of k priority; $\beta_k(s) = \int_0^\infty e^{-sx} dB_k(x)$ Laplace-Stieltjes transform of the $B_k(x)$ function;

$\beta_{k1}(x) = \int_0^\infty x dB_k(x)$ the moment of order 1 for $B_k(x)$; $\lambda_k$ - the parameter of the input flow and $\sigma_k = \lambda_1 + ... + \lambda_k$ where $k = 1,...,5$.

## 2. Literature Review
### 2.1. The case of the system $M|G|1$

Within this waiting system, we shall study the waiting time given by

### 2.1.1. Service in reverse order (LIFO)

$$w(s) = (1 - a\beta_1) + \frac{a(1 - \pi(s))}{s + a - a\pi(s)},$$

where the Laplace-Stieltjes transform of the distribution function of the busy period $\pi(s)$ is numerically determined by the functional Kendall equation $\pi(s) = \beta(s + a - a\pi(s))$.

The distribution function of the waiting time $W(x)$ is calculated by the numerical inversion of $w(x)$ by the Laplace-Stieltjes transform. Thus, we establish concrete values for the $W(x)$ function, by using several algorithms of numerical inversion. In the case of uniform and exponential distributions, to learn the parameters used in the modelling, we have applied the Pearson method, called the method of moments (see, for example monograph [1]). By using this method, we have learned the estimates for the distribution functions. These are:

The initial moment (empirical) of $k$ order, resulted from the formula: $v_k = \frac{1}{n}\sum_{i=1}^{n} X_i^k$,

where $X_1, X_2, ..., X_n$ is a selection of $n$ order, from the Poisson theoretical distribution with $a$ parameter

$$v_1 = \frac{1}{n}\sum_{i=1}^{n} X_i \quad (1)$$

In this case, the estimate is static and we can say that the estimate (1) is unbiased because the parameter of the input flow is given by:

$$M(v_1) = \frac{1}{n}\sum_{i=1}^{n} M(X_i) = a$$

**Observation**. The estimate (1) is sufficient because it converges in probability, towards the parameter $a$ of the law of large numbers (I. Chebîşev), resulting

$$P\{|v_1 - a| < \varepsilon\} \to 1 \text{ for } n \to \infty$$

To estimate the parameter of the input flow $a$, we have used, for the uniform distribution, the following formula: $a = \frac{1}{n}\sum_{i=1}^{n} X_i$, where $X_i(X_1, X_2, ..., X_n)$ are the times of arrival in the seaport of $n$ ships in a certain time period.

In the case of exponential distribution, we have used the formula:

$$\frac{1}{b} = \frac{1}{n}\sum_{i=1}^{n} X_i, \qquad (2)$$

where $X_i$ is the service time of the ship $i$. In this way we find the parameter $b$.

**2.1.2. If the service is in direct order (FIFO)**

$$w(s) = (1 - a\beta_1)\frac{s}{s - a + a\beta(s)}$$

**2.2. The case of the system $M_r | G_r | 1$ with continuation of the interrupted service**

In this case, the traffic coefficient is calculated as follows:

$$\rho_{k1} = \lambda_1\beta_{11} + \lambda_2\beta_{21} + ... + \lambda_k\beta_{k1}.$$

The system is viable if the traffic coefficient is less than 1.

Next, we shall analyse this traffic coefficient if the service time of ships in the seaport has exponential distribution, uniform in the period $[a_k, b_k]$, is an Erlang distribution of order 2 or a Gamma distribution with parameter $\alpha = 3$.

**2.3. The case of the system $M_r | G_r | 1$ with the loss of the interrupted message**

In this case, the traffic coefficient is calculated as follows:

$$\rho_{k1} = \lambda_1\beta_1 + \frac{\lambda_2}{\sigma_1}[1 - \beta_2(\sigma_1)] + ... + \frac{\lambda_k}{\sigma_{k-1}}[1 - \beta_k(\sigma_{k-1})],$$

where $\sigma_k = \lambda_1 + ... + \lambda_k$. The system is viable if the traffic coefficient is less than 1.

**2.4. The case of the system $M_r | G_r | 1$ when the interrupted message is served from the beginning**

In this case, the traffic coefficient is calculated as follows:

$$\rho_{k1} = \lambda_1\beta_1 + \frac{\lambda_2}{\sigma_1}\left[\frac{1}{\beta_2(\sigma_1)} - 1\right] + ... + \frac{\lambda_k}{\sigma_{k-1}}\left[\frac{1}{\beta_k(\sigma_{k-1})} - 1\right],$$

where $\sigma_k = \lambda_1 + ... + \lambda_k$. The system is viable if the traffic coefficient is less than 1.

## 3. Methodology

For developing these calculations in the mathematical model $M|G|1$, we have developed the programming algorithm in C++ with which we have calculated the waiting times and the inverses of these functions by the Laplace-Stieltjes transform.

The investigations made both within the Constanta Seaport, where we analysed the newsletters for the operation of vessels in the sea terminals during one month (February 2016) and the simulations made in the paper, we have found that the waiting time of a ship can be considerably reduced, the most appropriate being the exponential distribution.

## 4. Results
## 4.1. The case of the system $M|G|1$ when service is in reverse order (LIFO)

**4.1.1.** If the waiting time of messages is evenly distributed on the time period $[a^*, b]$, the distribution function $B(x) = \dfrac{x - a^*}{b - a^*}$ has the moment of order 1 $\beta_1 = M(x) = \dfrac{a^* + b}{2}$ and has the Laplace-Stieltjes transform $\beta(s) = \dfrac{1}{s(b - a^*)}(e^{-sa} - e^{-sb})$, $s > 0$

Table 4.1.1. Dependence on the parameter of input flow

| Current number | $s$ | $a^*$ | $b$ | $a$ | $x$ | $w(s)$ | $W(x)$ |
|---|---|---|---|---|---|---|---|
| 1 | 1 | 1 | 5 | 0,20 | 2 | 0,5577276 | 0,0582281 |
| 2 | 1 | 1 | 5 | 0,25 | 2 | 0,4405438 | 0,2048543 |
| 3 | 1 | 1 | 5 | 0,28 | 2 | 0,3691526 | 0,3234092 |
| 4 | 1 | 1 | 5 | 0,30 | 2 | 0,3211335 | 0,4218430 |
| 5 | 1 | 1 | 5 | 0,35 | 2 | 0,1996819 | 0,7911617 |

Table 4.1.2. Dependence on the parameter of uniform distribution.

| Current number | $s$ | $a^*$ | $b$ | $a$ | $x$ | $w(s)$ | $W(x)$ |
|---|---|---|---|---|---|---|---|
| 1 | 1 | 1 | 3 | 0,3 | 2 | 0,6122375 | 0,0041842 |
| 2 | 2 | 1 | 3 | 0,3 | 2 | 0,5279578 | 0,0910943 |
| 3 | 3 | 1 | 3 | 0,3 | 2 | 0,4904466 | 0,1364511 |
| 4 | 4 | 1 | 3 | 0,3 | 2 | 0,4696650 | 0,1637194 |
| 5 | 5 | 1 | 3 | 0,3 | 2 | 0,4565786 | 0,1817598 |

Table 4.1.3 Dependence on the parameter $s$ in determining the waiting time.

| Current number | $s$ | $a^*$ | $b$ | $a$ | $x$ | $w(s)$ | $W(x)$ |
|---|---|---|---|---|---|---|---|
| 1 | 1 | 1 | 3 | 0,3 | 2 | 0,6122375 | 0,0041842 |
| 2 | 2 | 1 | 3 | 0,3 | 2 | 0,5279578 | 0,0910943 |
| 3 | 3 | 1 | 3 | 0,3 | 2 | 0,4904466 | 0,1364511 |
| 4 | 4 | 1 | 3 | 0,3 | 2 | 0,4696650 | 0,1637194 |
| 5 | 5 | 1 | 3 | 0,3 | 2 | 0,4565786 | 0,1817598 |

**4.1.2.** If the waiting time of messages is exponentially distributed, then the distribution function $B(x) = 1 - e^{-bx}$ has the moment of order 1 $\beta_1 = M(x) = \frac{1}{b}$ and the Laplace-Stieltjes transform $\beta(s) = \frac{b}{s+b}$.

Table 4.1.4 Dependence on the parameter $b$ of the exponential distribution.

| Current number | $b$ | $s$ | $x$ | $a$ | $w(s)$ | $W(x)$ |
|---|---|---|---|---|---|---|
| 1 | 10 | 1 | 2 | 16 | 0,2783011 | 0,5275267 |
| 2 | 11 | 1 | 2 | 16 | 0,4116243 | 0,2495373 |
| 3 | 12 | 1 | 2 | 16 | 0,5190001 | 0,1015047 |
| 4 | 13 | 1 | 2 | 16 | 0,6059155 | 0,0100826 |
| 5 | 9 | 1 | 2 | 16 | 0,1111112 | 1,3303402 |

Table 4.1.5. Dependence on the parameter $s$ in determining the waiting time.

| Current number | $b$ | $s$ | $x$ | $a$ | $w(s)$ | $W(x)$ |
|---|---|---|---|---|---|---|
| 1 | 10 | 1 | 2 | 12 | 0,6000005 | 0,0156847 |
| 2 | 10 | 2 | 2 | 12 | 0,5101022 | 0,1120998 |
| 3 | 10 | 3 | 2 | 12 | 0,4479205 | 0,1940898 |
| 4 | 10 | 4 | 2 | 12 | 0,4000001 | 0,2687076 |
| 5 | 10 | 5 | 2 | 12 | 0,3610134 | 0,3388641 |

Table 4.1.6. Dependence on the parameter of input flow.

| Current number | $b$ | $s$ | $x$ | $a$ | $w(s)$ | $W(x)$ |
|---|---|---|---|---|---|---|
| 1 | 10 | 1 | 2 | 13 | 0,5258346 | 0,0935392 |
| 2 | 10 | 1 | 2 | 14 | 0,4468874 | 0,1955827 |
| 3 | 10 | 1 | 2 | 15 | 0,3641102 | 0,3329298 |
| 4 | 10 | 1 | 2 | 16 | 0,2783011 | 0,5275267 |
| 5 | 10 | 1 | 2 | 17 | 0,1900982 | 0,8325626 |

**4.2. The case of the system $M|G|1$ when service is in direct order (FIFO)**

**4.2.1.** If the waiting time of messages is evenly distributed on the time period $[a^*, b]$, by applying the same formulas indicated for LIFO, we get the following data:

Table 4.2.1. Dependence on the parameter of input flow

| Current number | s | a* | b | a | x | w2(s) | W2(x) |
|---|---|---|---|---|---|---|---|
| 1 | 1 | 1 | 5 | 0,17 | 2 | 0,5796426 | 0,0356066 |
| 2 | 1 | 1 | 5 | 0,19 | 2 | 0,5198547 | 0,1005006 |
| 3 | 1 | 1 | 5 | 0,20 | 2 | 0,4889634 | 0,1383438 |
| 4 | 1 | 1 | 5 | 0,23 | 2 | 0,3920251 | 0,2822969 |
| 5 | 1 | 1 | 5 | 0,25 | 2 | 0,3235947 | 0,4163370 |

Table 4.2.2. Dependence on the parameter of uniform distribution.

| Current number | s | a* | b | a | x | w2(s) | W2(x) |
|---|---|---|---|---|---|---|---|
| 1 | 1 | 2 | 5 | 0,20 | 2 | 0,3710239 | 0,3199197 |
| 2 | 1 | 3 | 5 | 0,20 | 2 | 0,2486619 | 0,6139755 |
| 3 | 1 | 1 | 4 | 0,20 | 2 | 0,6073089 | 0,0087747 |
| 4 | 1 | 1 | 6 | 0,20 | 2 | 0,3682717 | 0,3250601 |
| 5 | 1 | 2 | 6 | 0,20 | 2 | 0,2479412 | 0,6162432 |

Table 4.2.3. Dependence on the parameter $s$ in determining the waiting time.

| Current number | s | a* | b | a | x | w2(s) | W2(x) |
|---|---|---|---|---|---|---|---|
| 1 | 1 | 1 | 3 | 0,3 | 2 | 0,5349640 | 0,0831249 |
| 2 | 2 | 1 | 3 | 0,3 | 2 | 0,4678460 | 0,1661854 |
| 3 | 3 | 1 | 3 | 0,3 | 2 | 0,4440361 | 0,1997278 |
| 4 | 4 | 1 | 3 | 0,3 | 2 | 0,4323522 | 0,2170991 |
| 5 | 5 | 1 | 3 | 0,3 | 2 | 0,4255136 | 0,2275648 |

**4.2.2.** If the waiting time of messages is exponentially distributed, we get:

Table 4.2.4 Dependence on the parameter $s$ in determining the waiting time.

| Current number | b | s | x | a | w2(s) | W2(s) |
|---|---|---|---|---|---|---|
| 1 | 5 | 1 | 3 | 4 | 0,6000000 | 0,0156852 |
| 2 | 5 | 2 | 3 | 4 | 0,4666667 | 0,1677913 |
| 3 | 5 | 3 | 3 | 4 | 0,4000000 | 0,2687077 |
| 4 | 5 | 4 | 3 | 4 | 0,3600000 | 0,3408206 |
| 5 | 5 | 5 | 3 | 4 | 0,3333333 | 0,3950777 |

Table 4.2.5. Dependence on the parameter of input flow.

| Current number | b | s | x | a | w2(s) | W2(s) |
|---|---|---|---|---|---|---|
| 1 | 5 | 1 | 3 | 4,1 | 0,5684211 | 0,0470317 |
| 2 | 5 | 1 | 3 | 4,3 | 0,4941176 | 0,1318006 |
| 3 | 5 | 1 | 3 | 4,5 | 0,4000000 | 0,2687077 |
| 4 | 5 | 1 | 3 | 4,7 | 0,2769231 | 0,5312752 |
| 5 | 5 | 1 | 3 | 4,8 | 0,2000000 | 0,7898325 |

## 4.3. The case of the system $M_r | G_r | 1$ with continuation of the interrupted service

**Example 4.3.1**: If in the seaport, the time between two arrivals of ships has exponential distribution with parameters $\lambda_k$, $k = 1,...,5$ and the service time of the ships is an exponential distribution, then we can compute the traffic coefficient, where the distribution function $B_k(x) = 1 - e^{-b_k x}$ has the Laplace-Stieltjes transform $\beta_k(s) = \frac{b_k}{s + b_k}$ and the moment of order 1 is $\beta_{k1} = M(x) = \frac{1}{b_k}$.

Table 4.3.1. Sevice time is an exponential distribution

| Current number (k) | $b_k$ | $\lambda_k$ | $\beta_{k1}$ | $\rho_k$ |
|---|---|---|---|---|
| 1 | 7 | 0,3 | 0,14 | 0,04 |
| 2 | 3 | 0,2 | 0,33 | 0,10 |
| 3 | 4 | 0,4 | 0,25 | 0,20 |
| 4 | 2 | 0,5 | 0,4 | 0,40 |
| 5 | 5 | 0,8 | 0,2 | 0,56 |

**Example 4.3.2**: If in the seaport, the time between two arrivals of ships has exponential distribution with parameters $\lambda_k$, $k = 1,...,5$ and the service time of the ships is a uniform distribution within the given time period $[a_k, b_k]$, then we can compute the traffic coefficient, where the distribution function $B_k(x) = \frac{x - a_k}{b_k - a_k}$ has the Laplace-Stieltjes transform $\beta_k(s) = \frac{1}{s(b_k - a_k)}(e^{-sa_k} - e^{-sb_k})$, and the moment of order 1 is $\beta_{k1} = M(x) = \frac{a_k + b_k}{2}$.

Table 4.3.2. Service time is a uniform distribution

| Current number (k) | $[a_k, b_k]$ | $\lambda_k$ | $\beta_{k1}$ | $\rho_k$ |
|---|---|---|---|---|
| 1 | [1,4] | 0,3 | 2,5 | 0,75 |
| 2 | [2,5] | 0,2 | 3,5 | 1,45 |
| 3 | [1,6] | 0,4 | 3,5 | 2,85 |
| 4 | [4,7] | 0,5 | 5,5 | 5,6 |
| 5 | [1,7] | 0,8 | 4 | 8,8 |

**Example 4.3.3**: If in the seaport, the time between two arrivals of ships has exponential distribution with parameters $\lambda_k$, $k = 1,...,5$ and the service time of the ships is an Erlang distribution of order 2, then we can compute the traffic coefficient, where the Laplace-

Stieltjes transform of the function of distribution of the service time is: $\beta_k(s) = \left(\dfrac{b_k}{s+b_k}\right)^2$ and the moment of order 1 is $\beta_{k1} = M(x) = \dfrac{2}{b_k}$.

Table 4.3.3. Service time is an Erlang distribution

| Current number (k) | $b_k$ | $\lambda_k$ | $\beta_{k1}$ | $\rho_k$ |
|---|---|---|---|---|
| 1 | 7 | 0,3 | 0,28 | 0,08 |
| 2 | 3 | 0,2 | 0,66 | 0,21 |
| 3 | 4 | 0,4 | 0,5 | 0,41 |
| 4 | 2 | 0,5 | 1 | 0,91 |
| 5 | 5 | 0,8 | 0,4 | 1,23 |

**Example 4.3.4:** If in the seaport, the time between two arrivals of ships has exponential distribution with parameters $\lambda_k$, $k = 1,...,5$, and the service time of the ships is a Gamma distribution with $\alpha = 3$ parameter, then we can compute the traffic coefficient, where the Laplace-Stieltjes transform of the allocation function of the service time is:
$\beta_k(s) = \left(\dfrac{b_k}{s+b_k}\right)^3$ and when the moment of order 1 is $\beta_{k1} = M(x) = \dfrac{3}{b_k}$.

Table 4.3.4 Service time is a Gamma distribution

| Current number (k) | $b_k$ | $\lambda_k$ | $\beta_{k1}$ | $\rho_k$ |
|---|---|---|---|---|
| 1 | 7 | 0,3 | 0,67 | 0,20 |
| 2 | 3 | 0,2 | 0,42 | 0,28 |
| 3 | 4 | 0,4 | 0,51 | 0,48 |
| 4 | 2 | 0,5 | 0,29 | 0,63 |
| 5 | 5 | 0,8 | 0,57 | 1,09 |

**4.4. The case of the system $M_r \mid G_r \mid 1$ with the loss of the interrupted message**

Next, we shall analyse this traffic coefficient if the service time of ships from the seaport has exponential distribution, uniform in the time period $[a_k, b_k]$, is an Erlang distribution of order 2 or a Gamma distribution with $\alpha = 3$ parameter.

**Example 4.4.1**: If in the seaport, the time between two arrivals of ships has exponential distribution with parameters $\lambda_k$, $k = 1,...,5$ and the service time of ships is an exponential distribution, then we can compute the traffic coefficient, where the distribution function

$B_k(x) = 1 - e^{-b_k x}$ has the Laplace-Stieltjes transform $\beta_k(s) = \dfrac{b_k}{s + b_k}$ and the moment of order 1 is $\beta_1 = M(x) = \dfrac{1}{b_1} = 0,14$.

Table 4.4.1. Sevice time is an exponential distribution

| Current number (k) | $b_k$ | $\lambda_k$ | $\sigma_k$ | $\rho_k$ |
|---|---|---|---|---|
| 1 | 7 | 0,3 | 0,3 | 0,04 |
| 2 | 3 | 0,2 | 0,5 | 0,10 |
| 3 | 4 | 0,4 | 0,9 | 0,19 |
| 4 | 2 | 0,5 | 1,4 | 0,36 |
| 5 | 5 | 0,8 | 2,2 | 0,48 |

**Example 4.4.2**: If in the seaport, the time between two arrivals of ships has exponential distribution with parameters $\lambda_k$, $k = 1,...5$, and the service time of ships is a uniform distribution within the period given $[a_k, b_k]$, then we can compute the traffic coefficient, where the distribution function $B_k(x) = \dfrac{x - a_k}{b_k - a_k}$ has the Laplace-Stieltjes transform $\beta_k(s) = \dfrac{1}{s(b_k - a_k)}(e^{-sa_k} - e^{-sb_k})$, and the moment of order 1 is $\beta_1 = M(x) = \dfrac{a_1 + b_1}{2} = 3,5$.

Table 4.4.2. Service time is a uniform distribution

| Current number (k) | $[a_k, b_k]$ | $\lambda_k$ | $\sigma_k$ | $\rho_k$ |
|---|---|---|---|---|
| 1 | [2,5] | 0,3 | 0,3 | 1,05 |
| 2 | [2,7] | 0,2 | 0,5 | 1,53 |
| 3 | [1,3] | 0,4 | 0,9 | 2,02 |
| 4 | [3,8] | 0,5 | 1,4 | 2,57 |
| 5 | [1,8] | 0,8 | 2,2 | 3,13 |

**Example 4.4.3**: If in the seaport, the time between two arrivals of ships has exponential distribution with parameters $\lambda_k$, $k = 1,...5$, and the service time of the ships is an Erlang distribution of order 2, then we can compute the traffic coefficient, where the distribution function of the serving time has the Laplace-Stieltjes transform: $\beta_k(s) = \left(\dfrac{b_k}{s + b_k}\right)^2$ and the moment of order 1 is $\beta_1 = M(x) = \dfrac{2}{b_1} = 0,28$.

Table 4.4.3. Service time is an Erlang distribution

| Current number (k) | $b_k$ | $\lambda_k$ | $\sigma_k$ | $\rho_k$ |
|---|---|---|---|---|
| 1 | 7 | 0,3 | 0,3 | 0,08 |
| 2 | 3 | 0,2 | 0,5 | 0,2 |
| 3 | 4 | 0,4 | 0,9 | 0,36 |
| 4 | 2 | 0,5 | 1,4 | 0,66 |
| 5 | 5 | 0,8 | 2,2 | 0,88 |

**Example 4.4.4**: If in the seaport, the time between two arrivals of ships has an exponential distribution with parameters $\lambda_k$, $k = 1,...,5$, and the service time of the vessels is a Gamma distribution with $\alpha = 3$ parameter, then we can compute the traffic coefficient, where the distribution function of the service time has the Laplace –Stieltjes transform: $\beta_k(s) = \left(\dfrac{b_k}{s+b_k}\right)^3$ and the moment of order 1 is $\beta_1 = M(x) = \dfrac{3}{b_1} = 0,4$

Table 4.4.4 Service time is a Gamma distribution

| Current number (k) | $b_k$ | $\lambda_k$ | $\sigma_k$ | $\rho_k$ |
|---|---|---|---|---|
| 1 | 7 | 0,3 | 0,3 | 0,12 |
| 2 | 3 | 0,2 | 0,5 | 0,28 |
| 3 | 4 | 0,4 | 0,9 | 0,52 |
| 4 | 2 | 0,5 | 1,4 | 0,90 |
| 5 | 5 | 0,8 | 2,2 | 1,20 |

## 4.5. The case of the system $M_r|G_r|1$ when the interrupted message is served from the beginning

Next, we shall analyse this traffic coefficient if the service time of ships from the seaport has exponential distribution, uniform in the period $[a_k, b_k]$, is an Erlang distribution of order 2 or a Gamma distribution with $\alpha = 3$ parameter.

**Example 4.5.1**: If in the seaport, the time between two arrivals of ships has exponential distribution with parameters $\lambda_k$, $k = 1,...,5$, and the service time of the ships is an exponential distribution, then we can compute the traffic coefficient, where the distribution function $B_k(x) = 1 - e^{-b_k x}$ has the Laplace-Stieltjes transform $\beta_k(s) = \dfrac{b_k}{s+b_k}$ and the moment of order 1 is $\beta_1 = M(x) = \dfrac{1}{b_1} = 0,14$.

Table 4.5.1. Sevice time is an exponential distribution

| Current number (k) | $b_k$ | $\lambda_k$ | $\sigma_k$ | $\rho_k$ |
|---|---|---|---|---|
| 1 | 7 | 0,3 | 0,3 | 0,04 |
| 2 | 3 | 0,2 | 0,5 | 0,10 |
| 3 | 4 | 0,4 | 0,9 | 0,20 |
| 4 | 2 | 0,5 | 1,4 | 0,45 |
| 5 | 5 | 0,8 | 2,2 | 0,61 |

**Example 4.5.2**: If in the seaport, the time between two arrivals of ships has exponential distribution with parameters $\lambda_k$, $k = 1,...,5$, and the service time of the ships is a uniform distribution within the period given $[a_k, b_k]$, then we can compute the traffic, where the distribution function is the Laplace-Stieltjes transform $\beta_k(s) = \dfrac{1}{s(b_k - a_k)}(e^{-sa_k} - e^{-sb_k})$, and the moment of order 1 is $\beta_1 = M(x) = \dfrac{a_1 + b_1}{2} = 3,5$.

Table 4.5.2. Service time is a uniform distribution

| Current number (k) | $[a_k, b_k]$ | $\lambda_k$ | $\sigma_k$ | $\rho_k$ |
|---|---|---|---|---|
| 1 | [2,5] | 0,3 | 0,3 | 1,05 |
| 2 | [2,7] | 0,2 | 0,5 | 2,76 |
| 3 | [1,3] | 0,4 | 0,9 | 4,06 |
| 4 | [3,8] | 0,5 | 1,4 | 43,18 |
| 5 | [1,8] | 0,8 | 2,2 | 65,47 |

**Example 4.5.3**: If in the seaport, the time between two arrivals of ships has exponential distribution with parameters $\lambda_k$, $k = 1,...,5$, and the service time of the ships is an Erlang distribution of order 2, then we can compute the traffic coefficient, where the distribution function of the service time has the Laplace-Stieltjes transform: $\beta_k(s) = \left(\dfrac{b_k}{s + b_k}\right)^2$ and the moment of order 1 is $\beta_1 = M(x) = \dfrac{2}{b_1} = 0,28$.

Table 4.5.3. Service time is an Erlang distribution

| Current number (k) | $b_k$ | $\lambda_k$ | $\sigma_k$ | $\rho_k$ |
|---|---|---|---|---|
| 1 | 7 | 0,3 | 0,3 | 0,08 |
| 2 | 3 | 0,2 | 0,5 | 0,23 |
| 3 | 4 | 0,4 | 0,9 | 0,44 |
| 4 | 2 | 0,5 | 1,4 | 1,06 |
| 5 | 5 | 0,8 | 2,2 | 1,43 |

**Example 4.5.4**: If in the seaport, the time between two arrivals of ships has exponential distribution with the parameters $\lambda_k$, $k=1,...,5$, and the service time of ships is a Gamma distribution with $\alpha = 3$ parameter, then we can compute the traffic coefficient, where the distribution function of the service time has the Laplace –Stieltjes transform: $\beta_k(s) = \left(\dfrac{b_k}{s+b_k}\right)^3$ and the moment of order 1 is $\beta_1 = M(x) = \dfrac{3}{b_1} = 0,4$.

Table 4.5.4 Service time is a Gamma distribution

| Current number (k) | $b_k$ | $\lambda_k$ | $\sigma_k$ | $\rho_k$ |
|---|---|---|---|---|
| 1 | 7 | 0,3 | 0,3 | 0,12 |
| 2 | 3 | 0,2 | 0,5 | 0,35 |
| 3 | 4 | 0,4 | 0,9 | 2,05 |
| 4 | 2 | 0,5 | 1,4 | 3,23 |
| 5 | 5 | 0,8 | 2,2 | 3,87 |

## 5. Discussion

The numerical modelling of the traffic coefficient depending on the given initial characteristics of the maritime terminal is presented in tables 4.3.1 - 4.5.4. They are considered initial characteristics, the distribution functions of service with their numerical parameters and the parameters of the input flow for the given class. By varying these parameters, we can achieve values of $\rho_{k1}$ less than 1, thereby ensuring a normal working process without overloading the terminal. As seen from the tables presented, only the data presented in Table 4.3.1, Table 4.4.1, Table 4. 4.3 and Table 4.5.1 ensure a steady process without overload, because only the original data in these tables allow us to achieve that all are less than 1. Obviously, it is sufficient that a single value of the coefficient $\rho_{k1}$ to be greater than or equal to 1 (as in the case of tables 4.3.3, 4.3.4 and 4.4.4) in order for the service to be entirely stopped, not taking into account that in the remaining classes, the process is stationary, given the fact that $\rho_1,...,\rho_3$ are less than 1. The modelling also shows us the priority class where we must intervene in order to ensure the terminal operation without overloading.

## 6. Conclusions

According to the modelling developed in C ++, we recommend in the future to take into account the fact that the system is more viable in the case of the exponential distribution.